\documentclass[10pt,twocolumn,letterpaper]{article}
\usepackage{authblk}
\usepackage[pagenumbers]{cvpr} 

\usepackage{lipsum}
\usepackage{multirow}

\DeclareMathOperator*{\argmin}{arg\,min}

\definecolor{cvprblue}{rgb}{0.21,0.49,0.74}
\usepackage[pagebackref,breaklinks,colorlinks,allcolors=cvprblue]{hyperref}

\def\paperID{*****} 
\def\confName{CVPR}
\def\confYear{2025}

\title{Rethinking Few-Shot Medical Image Segmentation by SAM2: A Training-Free Framework with Augmentative Prompting and Dynamic Matching}

\author[a, \#]{Haiyue Zu}
\author[a, \#]{Jun Ge}
\author[a]{Heting Xiao}
\author[a]{Jile Xie}
\author[a]{Zhangzhe Zhou}
\author[b]{Yifan Meng}
\author[a]{Jiayi Ni}
\author[a]{Junjie Niu}
\author[a]{Linlin Zhang}
\author[a, $*$]{Li Ni}
\author[a, $*$]{Huilin Yang}
\affil[a]{Department of Orthopaedics, The First Affiliated Hospital of Soochow University, Soochow University, Suzhou, 215006, China.}
\affil[b]{Independent Researcher.}
\affil[ ]{\textsuperscript{\#}These authors contribute equally.}

\begin{document}
\maketitle

\begin{abstract}
The reliance on large labeled datasets presents a significant challenge in medical image segmentation. Few-shot learning offers a potential solution, but existing methods often still require substantial training data. This paper proposes a novel approach that leverages the Segment Anything Model 2 (SAM2), a vision foundation model with strong video segmentation capabilities. We conceptualize 3D medical image volumes as video sequences, departing from the traditional slice-by-slice paradigm. Our core innovation is a support-query matching strategy: we perform extensive data augmentation on a single labeled support image and, for each frame in the query volume, algorithmically select the most analogous augmented support image. This selected image, along with its corresponding mask, is used as a mask prompt, driving SAM2's video segmentation. This approach entirely avoids model retraining or parameter updates. We demonstrate state-of-the-art performance on benchmark few-shot medical image segmentation datasets, achieving significant improvements in accuracy and annotation efficiency. This plug-and-play method offers a powerful and generalizable solution for 3D medical image segmentation.
\end{abstract}
\section{Introduction}

\begin{figure}[t]
    \centering
    \includegraphics[width=0.8\linewidth]{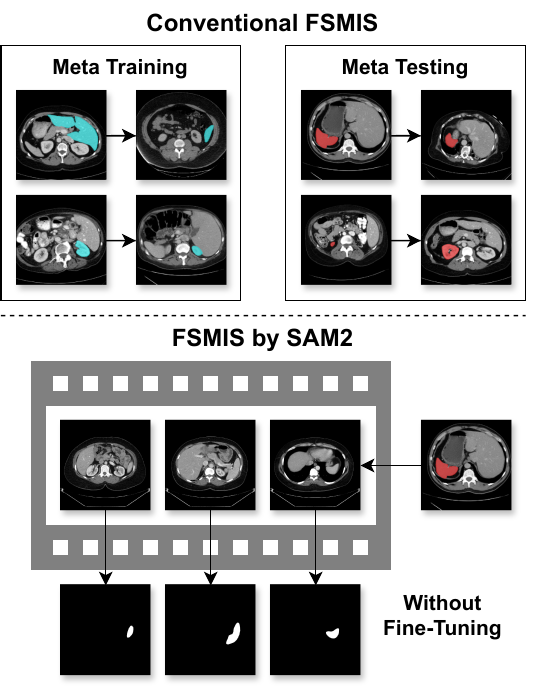}
    \caption{Visual comparison between conventional FSMIS methodologies and our novel paradigm. The former necessitates substantial meta-training using large annotated medical image repositories, whereas our framework capitalizes on the video segmentation prowess of SAM2 to eliminate training requirements, thereby significantly mitigating data labeling and computational training expenditures.}
    \label{fig:intro}
\end{figure}

The utilization of deep learning in medical image segmentation has become increasingly pivotal, facilitating accurate and automated analysis of medical images. However, traditional deep learning approaches suffer from several limitations, notably the reliance on large-scale labeled datasets and the requirement for training models from scratch, which contribute to exorbitant labeling and computational expenses.

To mitigate the annotation burden associated with traditional deep learning segmentation, few-shot segmentation has emerged as a promising paradigm. This approach leverages meta-learning principles to learn transferable knowledge from a set of base classes, enabling the model to rapidly adapt to unseen target classes with limited supervision. Typically, these methods involve a support set containing a few labeled examples of the target class, and a query image to be segmented. The model then utilizes the information from the support set to perform accurate segmentation on the query image. However, as Fig.~\ref{fig:intro} shows, most existing few-shot segmentation techniques still necessitate extensive labeled datasets for meta-training, indicating a substantial room for further improvement in reducing reliance on large-scale annotations.

The emergence of vision foundation models, such as the Segment Anything Model (SAM)~\cite{kirillov2023segment}, has revolutionized the field of image segmentation by demonstrating remarkable generalization capabilities across diverse downstream tasks, \textit{e.g.} medical imaging~\cite{mazurowski2023segment,huang2024segment} and remote sensing~\cite{osco2023segment,wang2024samrs}. SAM, trained on a vast corpus of data, has developed substantial segmentation expertise. The model's inherent ability to perform category-agnostic segmentation presents unique opportunities for few-shot learning, as it streamlines the adaptation of category-specific details. Therefore, the application of visual foundation models such as SAM in few-shot segmentation is a highly prospective research direction.

However, despite their impressive performance, achieving optimal results on specific downstream tasks like medical imageing often requires further adaptation. One common approach is interactive segmentation using prompts~\cite{cheng2023sam,wu2024one}, which is effective but still demands substantial time and expertise from clinicians, thereby limiting scalability and automation. Alternatively, fine-tuning SAM on task-specific datasets~\cite{wu2023medical,ma2024segment} can enhance performance but exacerbates the dependency on annotated data, counteracting the initial goal of reducing labeling efforts. Consequently, there is a critical need for a more efficient and tailored adaptation strategy that balances performance, scalability, and annotation efficiency.

Building upon the success of its predecessor, the latest iteration of the SAM, SAM2~\cite{ravi2024sam}, introduces groundbreaking advancements in video segmentation capabilities. Unlike SAM, which primarily focuses on static images, SAM2 can process video frame sequences by leveraging prompts added to one or a few frames to propagate segmentation across the entire sequence. This innovation is particularly promising for 3D medical imaging data, such as CT scans, which can be conceptualized as a series of 2D slices akin to a video stream. Furthermore, SAM2 enhances its versatility by introducing support for mask prompts, enabling users to provide deterministic segmentation labels for a single frame, which can then guide the segmentation of the entire sequence. This capability eliminates the ambiguity often associated with point or box prompts and reduces the reliance on interactive segmentation, which typically requires significant clinician involvement. Together, SAM2’s video segmentation and mask prompt functionalities create a powerful framework for automating medical image segmentation with minimal annotation effort.

We present a novel paradigm for few-shot medical image segmentation, departing from the conventional slice-by-slice approach as Fig.~\ref{fig:intro} depicts. We reframe the problem by considering 3D volumetric data as an integrated whole, leveraging SAM2's video segmentation capabilities to treat these volumes as sequences of 2D slices. Our core contribution lies in a support-query matching strategy: we perform extensive data augmentation on the support image and, for each frame in the query volume, algorithmically select the most analogous augmented support image. These optimal augmented support images, along with its mask, is then seamlessly inserted as a mask prompt into the query stream, driving SAM2's video segmentation. This method entirely avoids parameter updates or architectural modifications to SAM2, eliminating the need for labeled data and associated training expenditure. The plug-and-play nature of our technique makes it readily applicable to other video segmentation models. Experimental results demonstrate a significant performance improvement over prior state-of-the-art methods in few-shot medical image segmentation, highlighting the power of our approach.

We evaluated our method on three benchmark datasets: Synapse-CT, CHAOS-MRI, and CMR. Results demonstrate significant improvements over SOTA methods, with Dice score increases of \(1.50\%\), \(0.40\%\), and \(5.39\%\), highlighting the effectiveness of our approach in the few-shot setting.
Our contributions are as three folds:
\begin{itemize}
    \item We propose a fundamental shift in the approach to few-shot medical image segmentation by conceptualizing 3D volumetric data as a cohesive video sequence, enabling the application of video segmentation models to the entirety of scans such as CT and MRI.
    \item We introduce a novel prompting strategy that combines data augmentation of the support set with a per-frame optimal matching algorithm to identify and insert the most relevant support image and mask as a prompt for SAM2's video segmentation, eliminating the need for model retraining.
    \item We exceed SOTA performance on benchmark few-shot medical image segmentation tasks, demonstrating a significant improvement in segmentation accuracy and efficiency compared to existing methods.
\end{itemize}
\section{Related Works}

\subsection{Few-Shot Segmentation}

Few-shot Semantic Segmentation (FSS) methodologies offer a compelling alternative for segmentation tasks by enabling effective generalization from a restricted number of samples to previously unseen classes through the facilitation of interaction between support and query branches.

Early investigations, such as OSLSM~\cite{shaban2017one}, employed a two-branch network architecture to address the few-shot segmentation task, leveraging information and network parameters derived from the support branch to enhance the prediction of the query segmentation mask. Subsequently, SG-One~\cite{zhang2020sg} introduced masked average pooling to generate a prototype of the target class, building upon the foundational principles of prototypical networks~\cite{snell2017prototypical}. This approach has since been widely adopted as a baseline within the FSS domain. To comprehensively exploit query feature information, PANet~\cite{wang2019panet} developed a prototypical alignment network designed to perform reverse prediction of the support segmentation mask. PFENet~\cite{tian2020prior} further augmented query feature information by generating prior knowledge through high-level features and implementing a multi-scale fusion strategy.

Contemporary FSS methodologies can be broadly classified into three primary categories: prototypical network-based models~\cite{li2021adaptive,zhang2021self}, transformer-based models~\cite{zhang2021few,shi2022dense}, and conditional network-based models~\cite{shaban2017one,rakelly2018conditional}. These approaches predominantly focus on the generation of optimal prototypes to accurately represent category-specific information or on the establishment of interactive mechanisms between support and query features to improve the capture of query feature information. Notably, recent research endeavors~\cite{lang2022learning,lang2022beyond} have departed from conventional methodologies by exploring the FSS problem from a novel perspective, namely, the extraction of background information to facilitate query image segmentation.

\subsection{Few-Shot Medical Image Segmentation}

The acquisition, annotation, and dissemination of medical imagery are inherently constrained by elevated costs, labor-intensive annotation processes, and stringent ethical and legal frameworks, leading to a paucity of available data. Consequently, few-shot medical image segmentation (FSMIS) emerges as a promising paradigm for addressing these challenges. Within the established taxonomy of FSS, FSMIS models are broadly categorized into those predicated on prototypical networks and those employing two-branch interaction mechanisms.

Prototypical network-based models~\cite{tang2021recurrent,hansen2022anomaly,ouyang2022self,wang2022few} have been specifically developed to mitigate the substantial heterogeneity and diversity characteristic of medical images. For instance, SSL-ALPNet~\cite{ouyang2022self} introduces a network architecture that generates adaptive local prototypes and leverages supervoxels for self-supervised learning, resulting in significant performance enhancements over prior methodologies. ADNet~\cite{hansen2022anomaly}, building upon PANet~\cite{wang2019panet}, incorporates a method for predicting query segmentation masks using learned, fixed thresholds, thereby mitigating errors induced by complex backgrounds. Q-Net~\cite{shen2023q} extends ADNet by substituting fixed thresholds with learnable, adaptive thresholds to simultaneously predict query segmentation masks from dual-scale features and subsequently fuse the resulting predictions. Furthermore, it incorporates a prototype refinement module to refine the query prototype during the inference phase. The effective utilization of support branch information within the query branch remains a significant challenge in medical imaging contexts.

Unlike the aforementioned approaches, which primarily rely on variations of prototypical networks or two-branch interactions, our work explores a fundamentally different direction. We posit that vision foundation models, pre-trained on vast and diverse datasets, possess inherent representational capabilities that can be effectively transferred to the FSMIS task. Through this novel integration, our method achieves superior performance compared to existing prototypical and two-branch approaches, as evidenced by comprehensive experimental evaluations.

\subsection{SAM for Medical Images}

SAM has demonstrated promising performance in natural image segmentation, and its potential in medical imaging is increasingly being explored. However, due to significant domain differences between natural and medical images, SAM's direct application to medical tasks often yields suboptimal results. Adapting SAM to the unique characteristics of medical imaging remains a critical challenge. Some approaches focus on leveraging SAM's promptable segmentation capability, where user-provided prompts guide the model to segment specific regions of interest. While this interactive method can achieve high accuracy, it requires manual input from medical professionals, such as points or bounding boxes, which limits its scalability and practicality in clinical settings. Other methods attempt to fine-tune SAM on large-scale medical image segmentation datasets to bridge the domain gap. Although fine-tuning can improve performance, it introduces significant data and computational costs. 

Our method leverages the video segmentation capabilities of SAM2, enabling a few-shot segmentation paradigm that requires no fine-tuning. Additionally, it eliminates the need for interactive prompts, making the segmentation process fully automated and scalable. This is particularly advantageous in medical imaging, where automation can significantly enhance efficiency and reduce the workload on healthcare professionals.

\begin{figure*}[t]
    \centering
    \includegraphics[width=\linewidth]{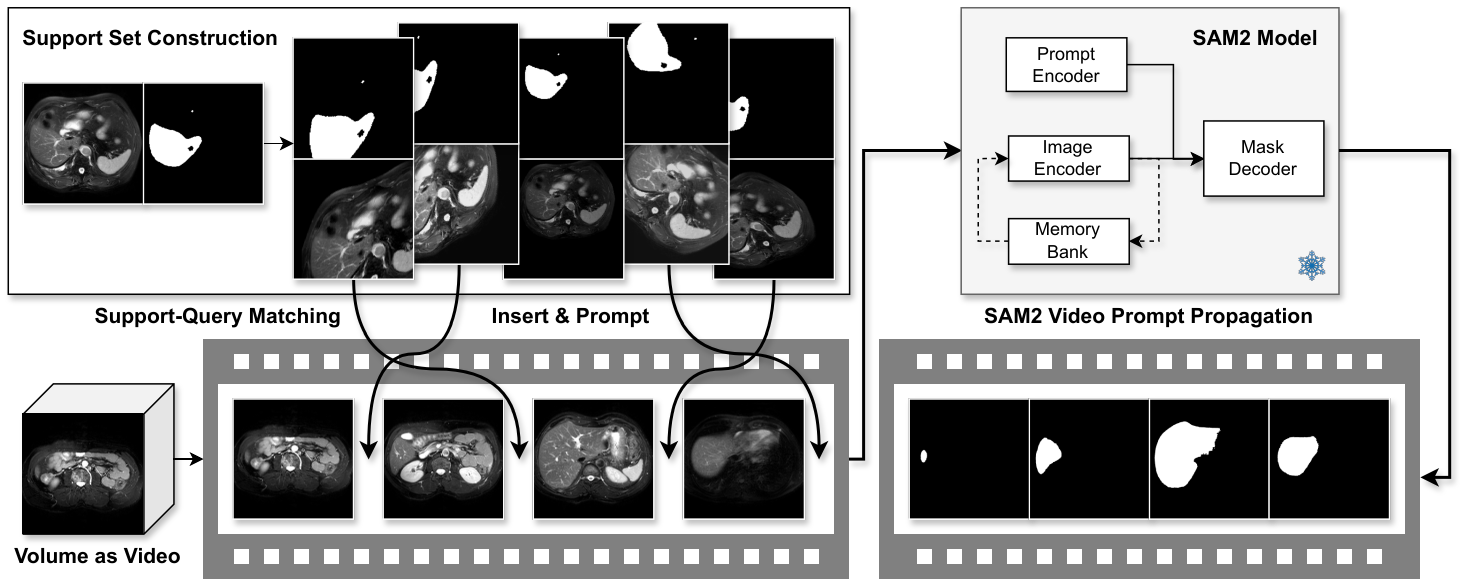}
    \caption{Overview of our proposed framework. The process involves: (1) Constructing a diverse support set through comprehensive augmentation of the single labeled support image; (2) Performing per-slice Support-Query Matching to identify the most perceptually similar augmented support image; (3) Utilizing SAM2 for Prompt-Driven Segmentation by treating the 3D query volume as a video sequence, with matched support images serving as mask prompts. This approach leverages SAM2's video segmentation capabilities without requiring any retraining.}
    \label{fig:overall_structure}
\end{figure*}
\section{Method}

\subsection{Problem Definition}

The FSMIS task is formalized as learning a model capable of segmenting unseen object classes $\mathcal C_{novel}$ after training on dataset $\mathcal D_{base}$, containing labeled samples from a disjoint set of base classes $\mathcal C_{base}$, where $\mathcal C_{base} \cap \mathcal C_{novel} = \varnothing$.  We adopt the episodic training paradigm prevalent in few-shot learning \cite{wang2019panet}. Each episode, representing an $N$-way $K$-shot learning task, is constructed by randomly sampling from a dataset $\mathcal D_{novel}$ containing novel classes.  This dataset is partitioned into a support set, $\mathcal S = \{(I_i^s, M_i^s)\}_{i=1}^{K}$, and a query set, $\mathcal Q = \{(I_j^q, M_j^q)\}_{i=1}^{N_q}$.  Here, $(I_i^s, M_i^s)$ represents the $i$-th image-mask pair in the support set, with $I_i^s$ denoting the image and $M_i^s$ the corresponding ground-truth segmentation mask.  Similarly, $(I_j^q, M_j^q)$ represents the $j$-th image-mask pair in the query set, where the ground truth mask $M^q$ is used only during training. The superscripts $j$ and $N_q$ denote the number of support and query samples per class, respectively.  The FSMIS model receives a support set $S$ and a query image $I^q$ as input and outputs a predicted binary segmentation mask, $\hat{M}^q$. Following prior work \cite{ouyang2022self,hansen2022anomaly}, we focus on the $1$-way $1$-shot setting.

\subsection{Architecture Overview}

Our architecture, presented in Fig.~\ref{fig:overall_structure}, introduces a novel approach to few-shot medical image segmentation, leveraging the power of SAM2's video segmentation in a three-stage process: Support Set Construction, Support-Query Matching, and Prompt-Driven Segmentation with SAM2. Unlike traditional methods, we first construct a highly diverse support set by applying a comprehensive suite of data augmentations to the single available support image, which maximizes the utility of the limited labeled data. Crucially, our Support-Query Matching stage introduces a per-slice matching algorithm. This algorithm selects, for each slice in the query volume, the most relevant augmented support image based on perceptual similarity. This optimal support image and its mask are then seamlessly integrated into the query sequence. The final stage leverages SAM2's inherent video segmentation capabilities. By treating the 3D query volume as a video sequence and using the matched support images as mask prompts, we guide SAM2 to perform accurate segmentation across the entire volume. This approach avoids any retraining or modification of SAM2, highlighting its plug-and-play nature and eliminating the need for extensive labeled data.

\subsection{Support Set Construction}

\begin{figure}[t]
    \centering
    \includegraphics[width=\linewidth]{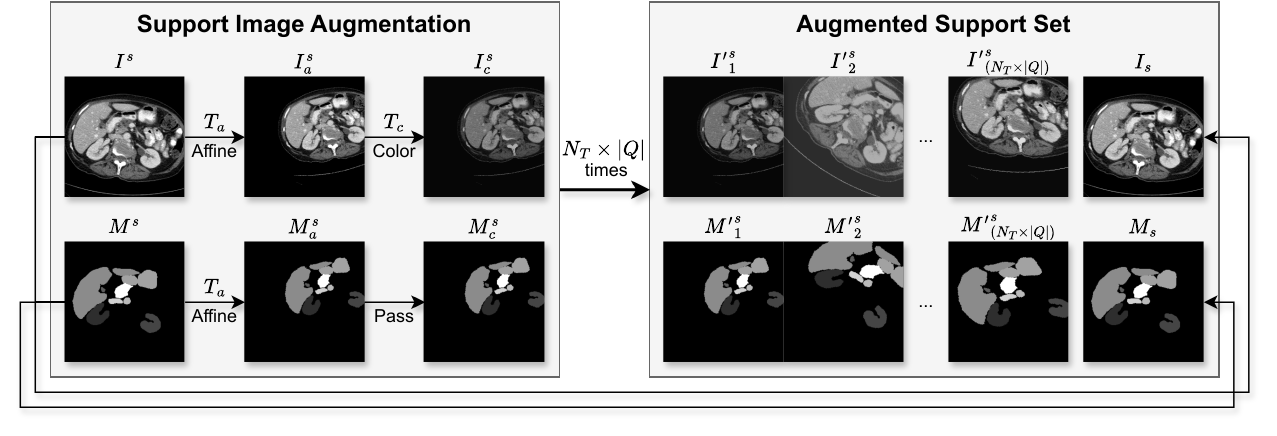}
    \caption{Schematic representation of the support set construction pipeline. This process involves the sequential application of affine transformations to both the support image and its corresponding mask, followed by color jittering applied solely to the image, thereby creating a diverse set of augmented support pairs.}
    \label{fig:support_set_construction}
\end{figure}

Given a single support image-mask pair, our goal is to generate a diverse set of augmented pairs to facilitate robust matching. We achieve this through a series of carefully designed geometric and photometric transformations as shown in Fig.~\ref{fig:support_set_construction}.

Let \( I^s \in \mathbb{R}^{H \times W \times C} \) represent the original support image, and \( M^s \in \mathbb N^{H \times W} \) be its corresponding binary mask, where \(H\), \(W\), and \(C\) denote the height, width, and number of channels, respectively.  We define a set of augmentation transformations, \( \mathcal{T} \), which includes both affine transformations and color jittering.

An affine transformation, \( T_a \in \mathcal{T}_a \), can be represented by a \( 3 \times 3 \) matrix:
\begin{equation}
 T_a = \begin{bmatrix}
 a_{11} & a_{12} & a_{13} \\
 a_{21} & a_{22} & a_{23} \\
 0 & 0 & 1
 \end{bmatrix},
\end{equation}
where \( a_{ij} \) are parameters controlling scaling, rotation, shearing, and translation.  Critically, the same transformation \( T_a \) is applied to both the support image \( I^s \) and its mask \( M^s \), preserving their correspondence:
\begin{equation}
\begin{aligned}
     I^s_a &= T_a(I^s) \\
 M^s_a &= T_a(M^s),
\end{aligned}
\end{equation}
where \( T_a(I^s) \) and \( T_a(M^s) \) represent the transformed image and mask, respectively. The transformation of the mask is achieved by applying \(T_a\) to the coordinates of each pixel in the mask.

The color jittering transformation, \( T_c \in \mathcal{T}_c \), adjusts the brightness, contrast, saturation, and hue of the image.  This can be represented as a function:
\begin{equation}
    I_c^s = T_c(I_a^s),
\end{equation}
where \(I^s_c\) represents the result of the color jittering transformation. Crucially, color jittering is only applied to the image, leaving the mask unchanged:
\begin{equation}
    M_c^s = M_a^s.
\end{equation}

We generate \(N_T\times N_q\) augmented image-mask pairs by randomly sampling transformations from \( \mathcal{T}_a \) and \( \mathcal{T}_c \), where \(N_T\) is defined as the number of support images per query image slice. Specifically, for each augmented pair \( (I^s_i, M^s_i) \), where \( i \in \{1, 2, ..., N_T\times N_q\} \):
\begin{equation}
\begin{aligned}
    T_a &\sim \mathcal{T}_a \\
  T_c &\sim \mathcal{T}_c \\
  {{I'}^s} &= T_c(T_a(I^s)) \\
  {{M'}^s} &= T_a(M^s).
\end{aligned}
\end{equation}

The hyperparameter \(N_T\) controls the size of the augmented support set Larger values of \(N_T\) increase the diversity of the support set, potentially leading to better matching, but also increase computational cost. The final augmented support set is then \( \mathcal{S}' =\bigcup^{K}_{j=1} \big(\{ I^s_j,M^s_j \} \cup \{({{I'}^s_i}, {{M'}^s_i})\}_{i=1}^{N_T\times N_q+1}\big) \), where the original support image and its mask are also included to prevent potential performance degradation.

\subsection{Support-Query Matching}

\begin{figure}[t]
    \centering
    \includegraphics[width=\linewidth]{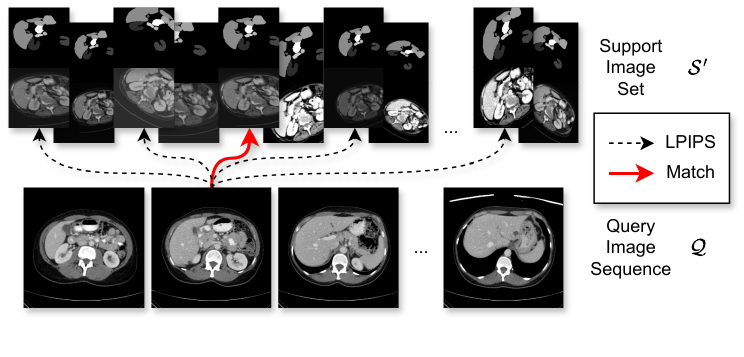}
    \caption{Illustration of the support-query matching process. For each query image \(I^q_j\), LPIPS is computed against all support images within the enhanced support set \(\mathcal S'\), and the support image \( {{{I'}^s_{i^*(j)}}} \) yielding the lowest LPIPS is identified as the best match.}
    \label{fig:support_query_matching}
\end{figure}

Given a query volume represented as a sequence of 2D slices, \(\mathcal Q= \{I^q_j\}_{j=1}^{N_q} \), where \( I^q_j \in \mathbb{R}^{H \times W \times C} \) is the \( j \)-th slice, and the augmented support set \( \mathcal{S}' = \{({{I'}_i^s},{{M'}_i^s})^{K\times(N_T\times N_q+1)}\}  \), our objective is to find the best-matching support image \( {{{I'}^s_{i^*(j)}}} \) for each query slice \( I^q_j \).  We achieve this by minimizing a perceptual dissimilarity metric of a matching pair \(({{I'}_{i}^s},I^q_j)\) as depicted in Fig.~\ref{fig:support_query_matching}.

We employ the Learned Perceptual Image Patch Similarity (LPIPS) metric~\cite{zhang2018unreasonable}, denoted as \( \mathcal{L}_{LPIPS} \), to quantify the perceptual difference between images. LPIPS is particularly well-suited for this task because it leverages the feature representations of a pre-trained deep convolutional neural network (typically AlexNet~\cite{krizhevsky2012imagenet} or VGG~\cite{simonyan2014very}), which are known to capture perceptually relevant features. Unlike pixel-space comparisons, LPIPS is more robust to minor misalignments and variations in texture, which are common in medical images.

Formally, let \( \Phi_l(I) \) represent the feature map extracted from the \( l \)-th layer of the pre-trained network for image \( I \).  The LPIPS distance between two images, \( I_1 \) and \( I_2 \), is computed as:
\begin{equation}
    \mathcal{L}_{LPIPS}(I_1, I_2) = \sum_l w_l \cdot \text{MSE}(\Phi_l(I_1), \Phi_l(I_2)),
\end{equation}
where \( w_l \) are learned weights that scale the contribution of each layer's feature map, and MSE denotes the mean squared error. These weights are learned on a dataset of human perceptual judgments, making LPIPS a perceptually meaningful distance metric.

For each query slice \( I_j^q \), we find the index \( i^* \) of the best-matching support image by minimizing the LPIPS distance:
\begin{equation}
    i^*(j) = \argmin_{i \in \{0, 1, ..., |\mathcal S|\}} \mathcal{L}_{LPIPS}(I_j^q, I^s_i).
\end{equation}

The corresponding best-matching support image \( {{{I'}^s_{i^*(j)}}} \)and its mask \( {{{M'}^s_{i^*(j)}}} \) are then used as the prompt for SAM2's video segmentation. We also investigate alternative similarity metrics, such as SSIM. A comprehensive comparison of these metrics is presented in the ablation study section.

\begin{figure}
    \centering
    \includegraphics[width=\linewidth]{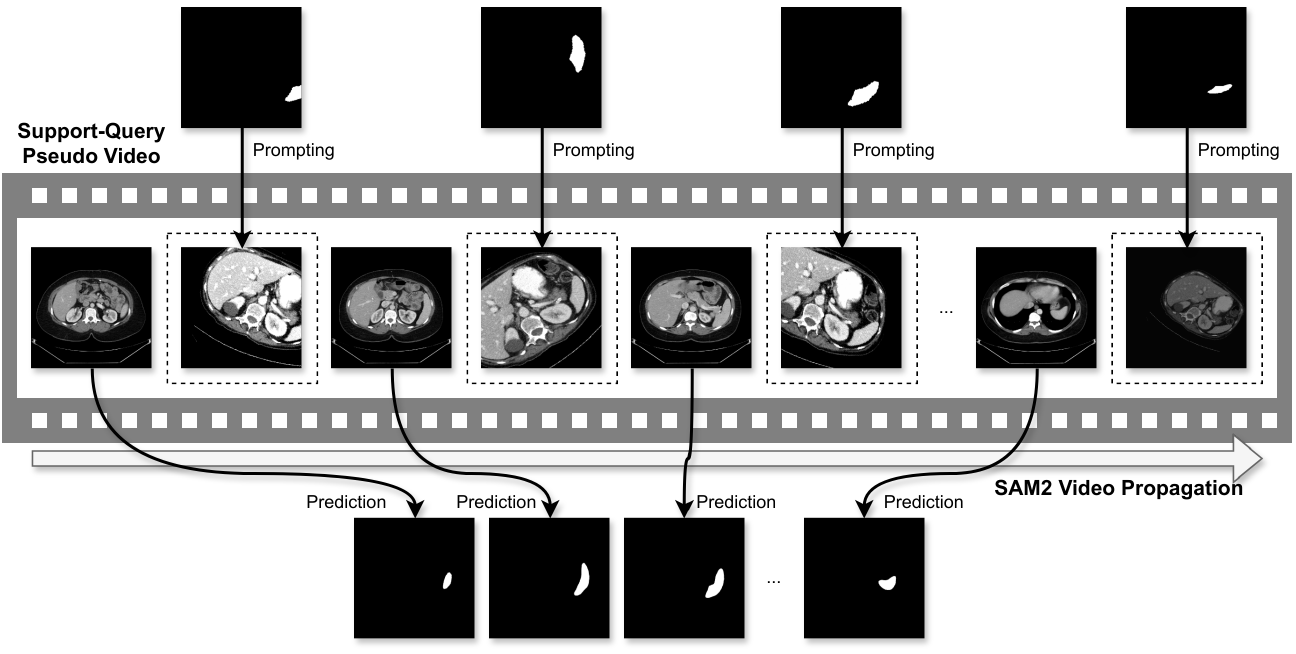}
    \caption{Illustration of the prompt-driven segmentation with SAM2. For each query slice, a two-frame input sequence is constructed, comprising the current query slice and the best-matching support image. The segmentation mask of the support image serves as a prompt for SAM2, enabling the propagation of segmentation to the query slice.}
    \label{fig:sam2_segmentation}
\end{figure}

\subsection{Prompt-Driven Segmentation with SAM2}

\subsubsection{Preliminary on SAM2}

SAM 2 represents an advanced promptable visual segmentation framework specifically engineered for comprehensive image and video processing tasks. Given an input sequence of frames or images \( \mathbf X=\{ \mathbf x_{t} \}^T_{t=1} \), accompanied by optional user prompts \( \mathbf P=\{ \mathbf p_{t} \}^T_{t=1} \), the model generates corresponding segmentation masks \( \mathbf Y=\{ \mathbf y_{t} \}^T_{t=1} \) for each frame \( \mathbf x_t \). The architectural design of SAM2 comprises several key components that synergistically contribute to its segmentation capabilities.

At its core, the model employs an image encoder $\mathcal E_{\text{image}}$ that transforms each input frame \( \mathbf x_t \) into a high-dimensional feature embedding \(\mathbf f_t = \mathcal E_{\text{image}}(\mathbf x_t) \), capturing essential visual characteristics. Complementing this, a specialized prompt encoder \(\mathcal E_p\) processes user-provided prompts \( \mathbf p_t \) to generate corresponding prompt embeddings \(\mathbf q_t = \mathcal E_p(\mathbf p_t)\), enabling interactive and context-aware segmentation. To enhance temporal consistency and leverage historical information in video sequences, the model incorporates a memory bank \(\mathcal M_t\) that stores the previous \(i\) historical embeddings \(\mathbf E_i\) from preceding frames.

The integration of these components is facilitated through a sophisticated memory attention mechanism \(\mathcal A\), which effectively combines the current frame's feature embedding \( \mathbf f_t \), historical memory embeddings \(\mathcal M_t\), and prompt embeddings \( \mathbf q_t \). This attention mechanism enables the model to maintain spatial-temporal coherence and adapt to dynamic visual content. Finally, a mask decoder \(\mathcal D\) processes these integrated features to predict the precise segmentation mask \(\mathbf y_t\) for the current frame. The segmentation process can be formulated as follows:
\begin{equation}
    \begin{aligned}
        \mathbf y_t&=\mathcal D(\mathcal A(\mathbf f_t,\mathcal M_t,\mathbf q_t)),~~~\text{for }t=1,2,\dots,T,\\
    \mathcal M_t&=\left\{ \mathbb E_i \mid i\in\{ \max(j,0) \}^{t-1}_{j=t-i-1} \right\}.
    \end{aligned}
\end{equation}

\subsubsection{Support-Query Pseudo Video Segmentation}

\begin{table*}[t]
\footnotesize
\centering
    \caption{Comparative quantitative analysis of methodological performance by Dice Score on the Synapse-CT and the CHAOS-MI datasets, with the best (in \textbf{bold font}) and the second best (\underline{underlined}) results highlighted. ``–'' means not reported.}
    \label{tab:comparison_abdomen}
    \begin{tabular}{c|ccccc|ccccc}
    \toprule
    \multirow{2}{*}{Method}& \multicolumn{5}{c|}{Synapse-CT} & \multicolumn{5}{c}{CHAOS-MRI}   \\ 
     & Spleen & Liver & LK & RK & Mean & Spleen & Liver & LK & RK & Mean \\ \midrule
    PA-Net~\cite{wang2019panet}  & 36.04 & 49.55 & 20.67 & 21.19 & 32.86 & 40.58 & 50.40 & 30.99 & 32.19 & 38.53     \\
    SE-Net~\cite{roy2020squeeze} & 43.66 & 35.42 & 24.42 & 12.51 & 29.00 & 47.30 & 29.02 & 45.78 & 47.96 & 42.51    \\
    SSL-ALPNet~\cite{ouyang2020self} & 70.96 & 78.29 & 72.36 & 71.81 & 73.35 & 72.18 & 76.10 & 81.92 & 85.18 & 78.84      \\
    ADNet~\cite{hansen2022anomaly} & 63.48 & 77.24 & 72.13 & \textbf{79.06} & 72.97 & 72.29 & \underline{82.11} & 73.68 & 85.80 & 78.51      \\
    AASDCL~\cite{wu2022dual} & 72.30 & 78.04 & 74.58 & 73.19 & 74.52 & \underline{76.24} & 72.33 & 80.37 & 86.11 & 78.76      \\
    SR\&CL~\cite{wang2022few} & 73.41 & 76.06 & 73.45 & 71.22 & 73.53 & 76.01 & 80.23 & 79.34 & 87.42 & 80.77    \\
    CRAPNet~\cite{ding2023few} & 70.37 & 75.41 & 74.69 & 74.18 & 73.66 & 74.32 & 76.46 & 81.95 & 86.42 & 79.79   \\
    Q-Net~\cite{shen2023q} & - & - & - & - & - & 75.99 & 81.74 & 78.36 & 87.98 & 81.02    \\
    CAT-Net~\cite{lin2023few} & 67.65 & 75.31 & 63.36 & 60.05 & 66.59 & 68.83 & 78.98 & 74.01 & 78.90 & 75.18     \\ 
    RPT~\cite{zhu2023few} & \underline{79.13} & \underline{82.57} & 77.05 & 72.58 & 77.83 & \textbf{76.37} & \textbf{82.86} & 80.72 & 89.82 & 82.44  \\ 
    GMRD~\cite{cheng2024few} & 78.31 & 79.60 & \textbf{81.70} & 74.46 & \underline{78.52} & 76.09 & 81.42 & \underline{83.96} & \underline{90.12} & \underline{82.90}  \\ 
    \textbf{Ours} & \textbf{83.73} & \textbf{83.12} & \underline{78.58} & \underline{74.65} & \textbf{80.02} & 75.57  & 76.26  & \textbf{89.15}  & \textbf{92.23}  & \textbf{83.30} \\  \bottomrule
    \end{tabular}
\end{table*}

\begin{table}[t]
\footnotesize
\centering
    \caption{Comparative quantitative analysis of methodological performance by Dice Score on the CMR dataset, with the best (in \textbf{bold font}) and the second best (\underline{underlined}) results highlighted. ``–'' means not reported.}
    \label{tab:comparison_cmr}
    \begin{tabular}{c|cccc}
    \toprule
    \multirow{2}{*}{Method}& \multicolumn{4}{c}{CMR}  \\ 
     & LV-BP & LV-MYO & RV & Mean \\ \midrule
    PA-Net~\cite{wang2019panet}  & 58.04 & 25.18 & 12.86 & 32.02   \\
    SE-Net~\cite{roy2020squeeze} & 72.77 & 44.76 & 57.13 & 58.20    \\
    SSL-ALPNet~\cite{ouyang2020self} & 83.99 & 66.74 & 79.96 & 76.90     \\
   SSL-ALPNet~\cite{ouyang2022self}  & 83.98 & \underline{67.68} & \underline{82.15} & 77.94     \\
   ADNet~\cite{hansen2022anomaly}  & 87.53 & 62.43 & 77.31 & 75.76      \\
   AASDCL~\cite{wu2022dual}  & 85.21 & 64.03 & 79.13 & 76.12\\
    SR\&CL~\cite{wang2022few} & 84.74 & 65.83 & 78.41 & 76.32 \\
   CRAPNet~\cite{ding2023few} & 83.02 & 65.48 & 78.27 & 75.59    \\
  Q-Net~\cite{shen2023q}  & \underline{90.25} & 65.92 & 78.19 & 78.15   \\ 
  GMRD~\cite{cheng2024few} &  90.00 & 67.04 & 80.29 & 79.11\\ 
    \textbf{Ours} & \textbf{91.43} & \textbf{75.82} & \textbf{86.26} & \textbf{84.50}  \\  \bottomrule
    \end{tabular}
\end{table}

The final stage leverages SAM2's inherent video segmentation capabilities to propagate the segmentation guided by the matched support images across the entire query volume \( \mathcal Q \) as Fig.~\ref{fig:sam2_segmentation} shows. For each slice \( I^q_j \), we create a two-frame input sequence for SAM2, denoted as \( \mathbf{I}_j \). This sequence consists of the best-matching support image \( {{I'}^s_{i^*(j)}} \) identified in the previous Support-Query Matching stage followed by the current query slice \( I_j^q \):
\begin{equation}
    \mathbf{I}_j = \left[  I_j^q ,{{I'}^s_{i^*(j)}}\right].
\end{equation}

The corresponding mask \( {{M'}^s_{i^*(j)}} \) of the selected support image serves as a mask prompt for SAM2.  We denote the SAM2 model as a function \( \text{SAM2}(\cdot, \cdot) \), which takes the input sequence and the mask prompt as arguments.  The predicted mask for the query slice \( I_j^q \) is then obtained as:
\begin{equation}
    \hat{M}^q_j = \text{SAM2}(\mathbf{I}_j, {{M'}^s_{i^*(j)}} ),
\end{equation}
where \( \hat{M}^q_j \in \{0, 1\}^{H \times W} \) is a binary mask.  This leverages SAM2's ability to maintain temporal consistency across frames, effectively propagating the segmentation indicated by the mask prompt to the subsequent query slice.

The procedure is repeated for each query slice, resulting in sequence of segmentation mask:
\begin{equation}
    \mathbf{M} =  \{\hat{M}^q_j\}_{j=1}^{N_q},
\end{equation}
where entire process is executed without any fine-tuning or modification of the pre-trained SAM2 model.
\section{Experiments}

\subsection{Datasets}

In this study, we undertook a rigorous assessment of the proposed method's performance, employing two publicly accessible medical image datasets: 

\textbf{The Synapse-CT dataset}~\cite{landman2015miccai}, originating from the MICCAI 2015 Multi-Atlas Abdomen Labeling Challenge, was utilized. This dataset encompasses 30 three-dimensional abdominal CT scans. From this corpus, we selected and analyzed four specific organs—namely, the left kidney, right kidney, liver, and spleen—for the purpose of performance evaluation. 

\textbf{The CHAOS-MRI dataset}~\cite{kavur2021chaos}, derived from the ISBI 2019 Combined Healthy Abdominal Organ Segmentation Challenge, served as a complementary dataset. This dataset comprises 20 three-dimensional T2-SPIR MRI scans, with each scan containing approximately 36 slices.  Consistent with the Synapse-CT dataset, we focused on the segmentation of four analogous categories: the left kidney, right kidney, liver, and spleen.

\textbf{The CMR dataset}~\cite{zhuang2018multivariate}, derived from the MICCAI 2019 Multi-Sequence Cardiac MRI Segmentation Challenge, encompasses 35 three-dimensional cardiac magnetic resonance imaging volumes. Each volume is axially discretized into approximately 13 slices, and contains segmentation masks delineating three distinct cardiac structures: the left ventricular blood pool (LV-BP), the left ventricular myocardium (LV-MYO), and the right ventricle (RV).

\subsection{Experimental Settings}

Current methodologies in FSMIS predominantly utilize the evaluation protocols detailed in~\cite{ouyang2022self} for assessing model performance.  These protocols are typically bifurcated into two settings: 

\textbf{Setting 1}, which stipulates a training regimen where instances of test categories may be present in the background of input training images, yet remain unannotated.  This scenario indicates that the classes designated as novel for evaluation are not entirely unseen by the model during training.

\textbf{Setting 2}, in contrast, necessitates the absolute absence of novel class instances from both the foreground and background of the training image corpus. This stringent condition ensures that all novel classes encountered in test images are genuinely agnostic.  In terms of anatomical grouping for dataset preparation, considering the consistent simultaneous visualization of bilateral kidneys on a single slice in CT and MRI scans, these are aggregated into a ``lower abdomen'' category.  Similarly, liver and spleen are grouped under the ``upper abdomen'' label.  

However, our method is inherently training-free, thus precluding its classification within either Setting 1 or Setting 2.  For the sake of fair comparison, we opted to benchmark against the best-reported performance of other methods under \textbf{Setting 1}.

\subsection{Implementation Details}

For the purposes of standardized evaluation and to facilitate direct comparison with established benchmarks, the present study conforms to the methodological framework detailed in~\cite{ouyang2020self}. The implementation meticulously replicates the preprocessing steps presented in the aforementioned work. The experimental results presented herein are obtained under a 1-way 1-shot learning configuration. 

To rigorously evaluate the efficacy of our proposed segmentation methodology, and to preclude the possibility of attributing superior performance solely to a strong backbone, we deliberately employed the ``tiny'' variant of the SAM2 model. Furthermore, the parameter \(N_T\) was set to \(2\). This value represents a carefully considered balance: it ensures sufficient diversity within the support set to capture variations in object appearance, while simultaneously avoiding excessive computational burden associated with a larger number of optimal matching calculations. Critically, our proposed method is entirely training-free. This characteristic eliminates the need for extensive hyperparameter tuning, which is often a laborious and computationally demanding process, thereby significantly reducing the overhead associated with optimizing model performance. This training-free nature contributes to the method's ease of deployment and generalizability.

\begin{figure}[t]
    \centering
    \includegraphics[width=1\linewidth]{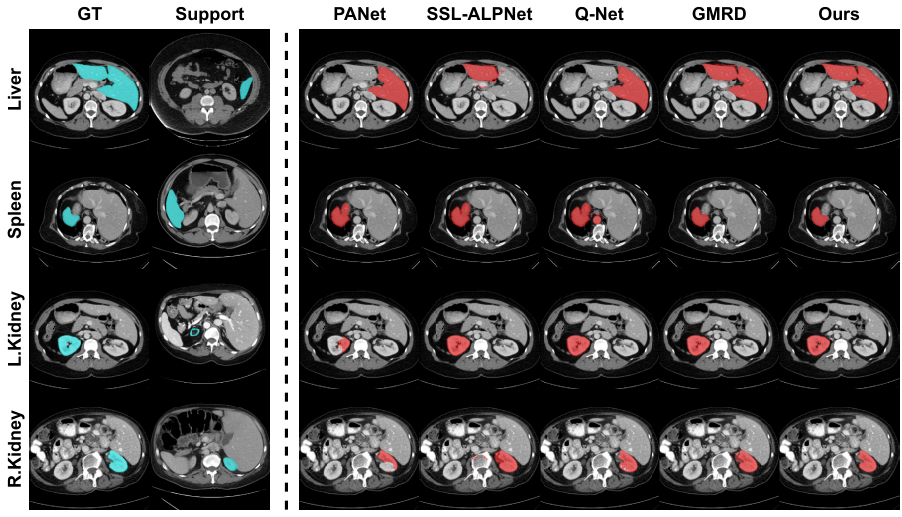}
    \caption{Visual comparison of segmentation results on the Synapse-CT dataset.}
    \label{fig:ct_comparison}
\end{figure}

\begin{figure}[t]
    \centering
    \includegraphics[width=1\linewidth]{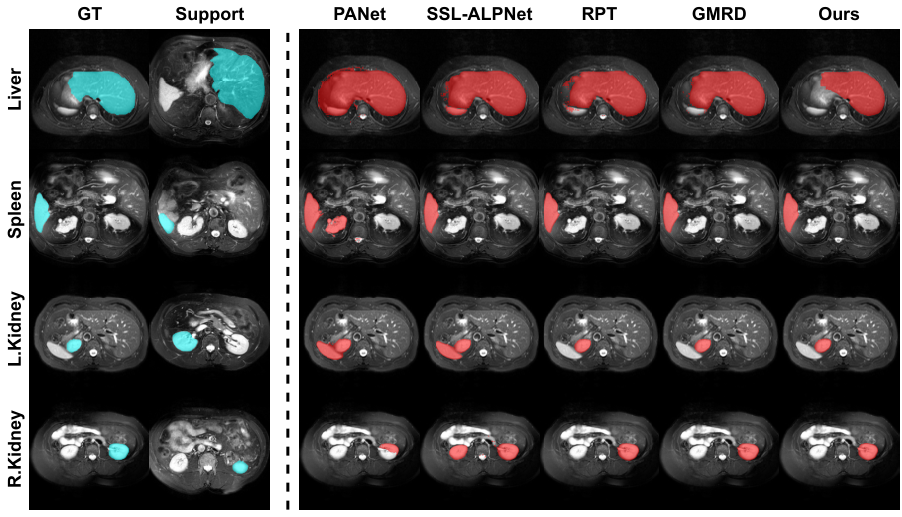}
    \caption{Visual comparison of segmentation results on the CHAOS-MRI dataset.}
    \label{fig:mri_comparison}
\end{figure}

\begin{figure}[t]
    \centering
    \includegraphics[width=1\linewidth]{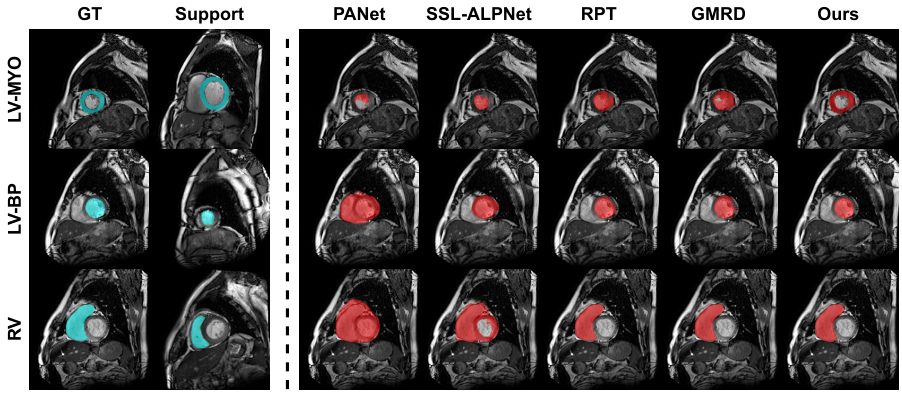}
    \caption{Visual comparison of segmentation results on the CMR dataset.}
    \label{fig:cmr_comparison}
\end{figure}

\subsection{Evaluation Metric}

To evaluate the proposed methodology, the Sorensen-Dice coefficient~\cite{ouyang2020self} was selected as the performance indicator, given its established utility in the FSMIS domain.  The Dice score, designed to assess the congruence between a model's predicted segmentation and the corresponding ground truth, is formally defined as:
\begin{equation}
    DSC(X,Y)=\frac{2| X\cap Y| }{|X|+|Y|}
\end{equation}
where \(X\) and \(Y\) are defined as the two sets being evaluated, and \(DSC\) represents the Dice Score.  The \(DSC\) value, ranging from \(0\) to \(1\), serves as a measure of similarity, with higher values indicative of greater resemblance between predictions and ground truth, and thus, superior segmentation performance.  The final experimental result constitutes the mean Dice score obtained through a 5-fold cross-validation protocol.

\subsection{Comparison With SOTA Methods}

To robustly demonstrate the improved efficacy and enhanced capability of our proposed approach, we performed a comparative evaluation against a set of representative and contemporary Few-Shot Medical Image Segmentation (FSMIS) methodologies, including SE-Net~\cite{roy2020squeeze}, PA-Net~\cite{wang2019panet}, SSL-ALPNet~\cite{ouyang2020self}, ADNet~\cite{hansen2022anomaly}, SR\&CL~\cite{wang2022few}, AASDCL~\cite{wu2022dual}, CRAPNet~\cite{ding2023few}, Q-Net~\cite{shen2023q}, CAT-Net~\cite{lin2023few}, RPT~\cite{zhu2023few} and GMRD~\cite{cheng2024few}. To thoroughly contextualize our findings, we provide a detailed overview of the experimental outcomes of current leading few-shot medical image segmentation models on four widely-used benchmark datasets, as presented in Tab.~\ref{tab:comparison_abdomen} and Tab.~\ref{tab:comparison_cmr}.

The results unequivocally demonstrate the superior performance of our method over all other evaluated methods without introducing any training cost.  Our method yielded a compelling mean Dice score of \(80.02\%\) on the Synapse-CT dataset, significantly outperforming the previous SOTA GRMD~\cite{cheng2024few} by \(1.50\%\).  Particularly striking is the performance in the Spleen and Liver region, where our method surpasses prior maximum scores by a considerable \(5.42\%\) and \(3.52\%\).  Similarly impressive results were achieved on the CHAOS-MRI dataset, with a test Dice score of \(83.30\%\), exceeding the current highest result by \(0.40\%\).  The gains are particularly pronounced in segmenting the left and right kidneys, achieving Dice scores \(5.19\%\) and \(2.11\%\) higher than previously reported benchmarks. Furthermore, our method demonstrated a substantial improvement on the CMR dataset, surpassing the current SOTA by \(5.39\%\).

\subsection{Visualization}

\begin{figure}[t]
    \centering
    \includegraphics[width=0.9\linewidth]{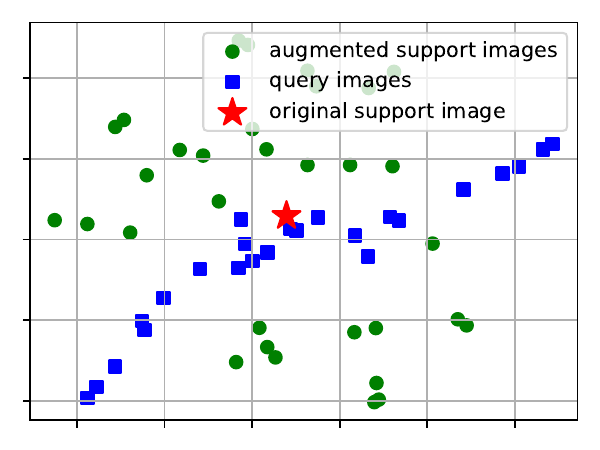}
    \caption{Visualization of the features of the original support image, augmented support images, and query images.}
    \label{fig:features}
\end{figure}

We provide visual comparisons of segmentation results in Fig.~\ref{fig:ct_comparison}, Fig.~\ref{fig:mri_comparison} and Fig.~\ref{fig:cmr_comparison}. Fig~\ref{fig:ct_comparison} and Fig.~\ref{fig:mri_comparison} showcase performance on the Synapse-CT and CHAOS-MRI datasets and Fig~\ref{fig:cmr_comparison} demonstrates results on the CMR dataset. In each case, our method demonstrates a clear advantage over the compared methods. The generated segmentation masks adhere more closely to the ground truth, exhibiting smoother contours, fewer spurious regions, and a more precise capture of the target shape. This superior performance is consistent across different modalities and anatomical targets.

To further investigate the mechanism behind the improved performance, we visualize the feature representations of the original support image, its augmented counterparts, and the query images in Fig.~\ref{fig:features}. As can be seen, the original support image shows strong feature similarity with only a limited portion of the query image sequence. However, the augmented support images exhibit a richer and more diverse feature set, facilitating more robust matching with a broader range of instances within the query sequence. This enhanced matching capability likely contributes to the observed improvements in segmentation accuracy.

\subsection{Ablation Studies}

To rigorously assess the contribution of individual model components and to quantify their impact on overall performance, a series of ablation studies are undertaken. These studies adhere to a controlled experimental design, manipulating one variable at a time while maintaining all other parameters constant, unless otherwise noted. The CHAOS-MRI dataset serves as the benchmark for these evaluations.

\subsubsection{Model Sizes}

\begin{table}[t]
    \footnotesize
    \centering
    \caption{Performance and parameter size comparison of different SAM2 variants.}
    \label{tab:model_sizes}
    \begin{tabular}{c|c|ccccc}
    \toprule
    \multirow{2}{*}{Model}& \multirow{2}{*}{Size (M)}& \multicolumn{5}{c}{CHAOS-MRI} \\ 
      &     &  Spleen & Liver & LK & RK & Mean \\  \midrule
     tiny & 38.9 & 75.57  & 76.26  & \textbf{89.15}  & 92.23  & 83.30 \\ 
     small & 46 & 77.24 & 77.19 & 88.82 & \textbf{92.41} & 83.91 \\ 
     base & 80.8 & 78.91 & 79.35 & 86.10 & 92.18 & 84.13 \\ 
     large & 224.4 & \textbf{79.31} & \textbf{79.51} & 86.28 & 91.94 & \textbf{84.26} \\ \bottomrule
    \end{tabular}
\end{table}

SAM2 offers a range of model sizes, from tiny to large, with increasing segmentation capability at the cost of higher computational demands. Tab~\ref{tab:model_sizes} presents a comparison of segmentation performance and parameter count across these different SAM variants. Notably, our method achieves superior performance compared to all baseline approaches even when utilizing the smallest tiny variant. Furthermore, we observe a consistent improvement in our method's performance as the SAM model size increases, demonstrating the scalability of our approach.

\subsubsection{Impact of support set construction}

\begin{table}[t]
    \footnotesize
    \centering
    \caption{Performance comparison of different \(N_T\).}
    \label{tab:n_t}
    \begin{tabular}{c|ccccc}
    \toprule
    \multirow{2}{*}{\(N_T\)} & \multicolumn{5}{c}{CHAOS-MRI} \\ 
        &  Spleen & Liver & LK & RK & Mean \\  \midrule
     0 & 68.11 & 75.49 & 85.34 & 83.35 & 78.07 \\ 
     1 & 75.05 & 76.10 & 87.85 & 90.50 & 82.37 \\ 
     2 & 75.57  & 76.26  & 89.15  & 92.23  & 83.30 \\ 
     4 & \textbf{76.51} & \textbf{77.15} & \textbf{89.31} & \textbf{92.96} & \textbf{83.98} \\ \bottomrule
    \end{tabular}
\end{table}

To maximize the utility of the limited labeled support images, we perform a series of augmentation operations on the support images and their corresponding masks, preserving pixel-level correspondences. This results in an augmented support set, denoted as  \(\mathcal{S}'\), with a size expanded to \(N_T \times N_Q + 1\). Here, \(N_T\) represents the number of augmented support images generated for each query image. A larger \(N_T\) leads to increased diversity within \(\mathcal{S}'\), potentially facilitating more robust support-query matching. However, this increased diversity comes at the cost of higher computational overhead.

To investigate the optimal balance between diversity and computational cost, we evaluate our method with different values of \(N_T \in\{0,1,2,4\}\). The case \(N_T = 0\) corresponds to the baseline scenario where no augmentation is applied, and only the original support images and masks are used. Our experimental results presented in Tab~\ref{tab:n_t} demonstrate that our method achieves strong performance at \(N_T = 2\), outperforming the baseline (\(N_T = 0\)). This significant improvement demonstrates the effectiveness of our proposed augmentation strategy in enhancing the representational capacity of the support set. While further increasing \(N_T\) may be tested, the performance in \(N_T=2\) already suggests a suitable trade-off.

\subsubsection{Selection of similarity metrics}

\begin{table}[t]
    \footnotesize
    \centering
    \caption{Performance comparison of different similarity metrics.}
    \label{tab:similarity_metric}
    \begin{tabular}{c|ccccc}
    \toprule
    \multirow{2}{*}{Metric} & \multicolumn{5}{c}{CHAOS-MRI} \\ 
        &  Spleen & Liver & LK & RK & Mean \\  \midrule
     LPIPS~\cite{zhang2018unreasonable} & \textbf{75.57}  & 76.26  & \textbf{89.15}  & \textbf{92.23}  & \textbf{83.30} \\ 
     SSIM~\cite{brunet2011mathematical} & 73.83 & \textbf{78.23} & 87.96 & 90.55 & 82.64 \\ 
     PSNR~\cite{korhonen2012peak} & 74.09 & 74.20 & 86.32 & 93.29 & 81.98\\ \bottomrule
    \end{tabular}
\end{table}

A core component of our proposed method involves selecting the optimal support image for a given query image based on a similarity metric. Our primary approach, as presented in the preceding sections, utilizes the LPIPS metric~\cite{zhang2018unreasonable} to quantify the perceptual distance between images. This choice is motivated by LPIPS's demonstrated ability to correlate well with human judgments of image similarity.

However, to assess the sensitivity of our method to the specific choice of similarity metric, we conducted an ablation study comparing LPIPS with other commonly used image similarity measures. Specifically, we considered the following alternative metrics: Structural Similarity Index (SSIM)~\cite{brunet2011mathematical} which measures the structural similarity between two images, considering luminance, contrast, and structure, and Peak Signal-to-Noise Ratio (PSNR)~\cite{korhonen2012peak} which is a classic metric that quantifies the ratio between the maximum possible power of a signal and the power of corrupting noise that affects the fidelity of its representation.

The results of this comparison are presented in Tab.~\ref{tab:similarity_metric}. As can be observed, the LPIPS metric achieved the best overall performance. This outcome reinforces the rationale for selecting a perceptually-aligned metric for optimal support image selection. Importantly, while LPIPS yielded the highest performance, the use of SSIM and PSNR did not result in a catastrophic performance drop. The performance degradation was relatively modest. This observation demonstrates the robustness of our proposed framework to variations in the similarity metric. The underlying mechanism of our method is resilient to the specific nuances of different similarity measures, provided they capture a reasonable notion of image similarity.

\section{Conclusion}

This paper presents a fundamentally new approach to few-shot medical image segmentation, challenging the conventional slice-by-slice processing of 3D volumetric data. By conceptualizing these volumes as video sequences, we unlock the power of SAM2's video segmentation capabilities. Our innovative support-query matching strategy, coupled with data augmentation, provides a highly effective prompting mechanism that obviates the need for any parameter updates to the foundation model. The substantial performance gains achieved on multiple benchmark datasets underscore the transformative potential of this approach. This work not only advances the SOTA in few-shot medical image segmentation but also demonstrates a generalizable strategy for leveraging foundation models without costly fine-tuning, offering a significant contribution to the broader field of medical image analysis.

{
    \small
    \bibliographystyle{ieeenat_fullname}
    \bibliography{main}

\begin{thebibliography}{42}
\providecommand{\natexlab}[1]{#1}
\providecommand{\url}[1]{\texttt{#1}}
\expandafter\ifx\csname urlstyle\endcsname\relax
  \providecommand{\doi}[1]{doi: #1}\else
  \providecommand{\doi}{doi: \begingroup \urlstyle{rm}\Url}\fi

\bibitem[Brunet et~al.(2011)Brunet, Vrscay, and Wang]{brunet2011mathematical}
Dominique Brunet, Edward~R Vrscay, and Zhou Wang.
\newblock On the mathematical properties of the structural similarity index.
\newblock \emph{IEEE Transactions on Image Processing}, 21\penalty0 (4):\penalty0 1488--1499, 2011.

\bibitem[Cheng et~al.(2023)Cheng, Qin, Jiang, Zhang, Lao, and Li]{cheng2023sam}
Dongjie Cheng, Ziyuan Qin, Zekun Jiang, Shaoting Zhang, Qicheng Lao, and Kang Li.
\newblock Sam on medical images: A comprehensive study on three prompt modes.
\newblock \emph{arXiv preprint arXiv:2305.00035}, 2023.

\bibitem[Cheng et~al.(2024)Cheng, Wang, Xin, Zhou, Zhang, and Shao]{cheng2024few}
Ziming Cheng, Shidong Wang, Tong Xin, Tao Zhou, Haofeng Zhang, and Ling Shao.
\newblock Few-shot medical image segmentation via generating multiple representative descriptors.
\newblock \emph{IEEE Transactions on Medical Imaging}, 43\penalty0 (6):\penalty0 2202--2214, 2024.

\bibitem[Ding et~al.(2023)Ding, Sun, Tang, Cai, and Yan]{ding2023few}
Hao Ding, Changchang Sun, Hao Tang, Dawen Cai, and Yan Yan.
\newblock Few-shot medical image segmentation with cycle-resemblance attention.
\newblock In \emph{Proceedings of the IEEE/CVF winter conference on applications of computer vision}, pages 2488--2497, 2023.

\bibitem[Hansen et~al.(2022)Hansen, Gautam, Jenssen, and Kampffmeyer]{hansen2022anomaly}
Stine Hansen, Srishti Gautam, Robert Jenssen, and Michael Kampffmeyer.
\newblock Anomaly detection-inspired few-shot medical image segmentation through self-supervision with supervoxels.
\newblock \emph{Medical Image Analysis}, 78:\penalty0 102385, 2022.

\bibitem[Huang et~al.(2024)Huang, Yang, Liu, Zhou, Chang, Zhou, Chen, Yu, Chen, Chen, et~al.]{huang2024segment}
Yuhao Huang, Xin Yang, Lian Liu, Han Zhou, Ao Chang, Xinrui Zhou, Rusi Chen, Junxuan Yu, Jiongquan Chen, Chaoyu Chen, et~al.
\newblock Segment anything model for medical images?
\newblock \emph{Medical Image Analysis}, 92:\penalty0 103061, 2024.

\bibitem[Kavur et~al.(2021)Kavur, Gezer, Bar{\i}{\c{s}}, Aslan, Conze, Groza, Pham, Chatterjee, Ernst, {\"O}zkan, et~al.]{kavur2021chaos}
A~Emre Kavur, N~Sinem Gezer, Mustafa Bar{\i}{\c{s}}, Sinem Aslan, Pierre-Henri Conze, Vladimir Groza, Duc~Duy Pham, Soumick Chatterjee, Philipp Ernst, Sava{\c{s}} {\"O}zkan, et~al.
\newblock Chaos challenge-combined (ct-mr) healthy abdominal organ segmentation.
\newblock \emph{Medical image analysis}, 69:\penalty0 101950, 2021.

\bibitem[Kirillov et~al.(2023)Kirillov, Mintun, Ravi, Mao, Rolland, Gustafson, Xiao, Whitehead, Berg, Lo, et~al.]{kirillov2023segment}
Alexander Kirillov, Eric Mintun, Nikhila Ravi, Hanzi Mao, Chloe Rolland, Laura Gustafson, Tete Xiao, Spencer Whitehead, Alexander~C Berg, Wan-Yen Lo, et~al.
\newblock Segment anything.
\newblock In \emph{Proceedings of the IEEE/CVF International Conference on Computer Vision}, pages 4015--4026, 2023.

\bibitem[Korhonen and You(2012)]{korhonen2012peak}
Jari Korhonen and Junyong You.
\newblock Peak signal-to-noise ratio revisited: Is simple beautiful?
\newblock In \emph{2012 Fourth international workshop on quality of multimedia experience}, pages 37--38. IEEE, 2012.

\bibitem[Krizhevsky et~al.(2012)Krizhevsky, Sutskever, and Hinton]{krizhevsky2012imagenet}
Alex Krizhevsky, Ilya Sutskever, and Geoffrey~E Hinton.
\newblock Imagenet classification with deep convolutional neural networks.
\newblock \emph{Advances in neural information processing systems}, 25, 2012.

\bibitem[Landman et~al.(2015)Landman, Xu, Igelsias, Styner, Langerak, and Klein]{landman2015miccai}
Bennett Landman, Zhoubing Xu, Juan Igelsias, Martin Styner, Thomas Langerak, and Arno Klein.
\newblock Miccai multi-atlas labeling beyond the cranial vault--workshop and challenge.
\newblock In \emph{Proc. MICCAI multi-atlas labeling beyond cranial vault—workshop challenge}, page~12. Munich, Germany, 2015.

\bibitem[Lang et~al.(2022{\natexlab{a}})Lang, Cheng, Tu, and Han]{lang2022learning}
Chunbo Lang, Gong Cheng, Binfei Tu, and Junwei Han.
\newblock Learning what not to segment: A new perspective on few-shot segmentation.
\newblock In \emph{Proceedings of the IEEE/CVF conference on computer vision and pattern recognition}, pages 8057--8067, 2022{\natexlab{a}}.

\bibitem[Lang et~al.(2022{\natexlab{b}})Lang, Tu, Cheng, and Han]{lang2022beyond}
Chunbo Lang, Binfei Tu, Gong Cheng, and Junwei Han.
\newblock Beyond the prototype: Divide-and-conquer proxies for few-shot segmentation.
\newblock \emph{arXiv preprint arXiv:2204.09903}, 2022{\natexlab{b}}.

\bibitem[Li et~al.(2021)Li, Jampani, Sevilla-Lara, Sun, Kim, and Kim]{li2021adaptive}
Gen Li, Varun Jampani, Laura Sevilla-Lara, Deqing Sun, Jonghyun Kim, and Joongkyu Kim.
\newblock Adaptive prototype learning and allocation for few-shot segmentation.
\newblock In \emph{Proceedings of the IEEE/CVF conference on computer vision and pattern recognition}, pages 8334--8343, 2021.

\bibitem[Lin et~al.(2023)Lin, Chen, Cheng, and Chen]{lin2023few}
Yi Lin, Yufan Chen, Kwang-Ting Cheng, and Hao Chen.
\newblock Few shot medical image segmentation with cross attention transformer.
\newblock In \emph{International Conference on Medical Image Computing and Computer-Assisted Intervention}, pages 233--243. Springer, 2023.

\bibitem[Ma et~al.(2024)Ma, He, Li, Han, You, and Wang]{ma2024segment}
Jun Ma, Yuting He, Feifei Li, Lin Han, Chenyu You, and Bo Wang.
\newblock Segment anything in medical images.
\newblock \emph{Nature Communications}, 15\penalty0 (1):\penalty0 654, 2024.

\bibitem[Mazurowski et~al.(2023)Mazurowski, Dong, Gu, Yang, Konz, and Zhang]{mazurowski2023segment}
Maciej~A Mazurowski, Haoyu Dong, Hanxue Gu, Jichen Yang, Nicholas Konz, and Yixin Zhang.
\newblock Segment anything model for medical image analysis: an experimental study.
\newblock \emph{Medical Image Analysis}, 89:\penalty0 102918, 2023.

\bibitem[Osco et~al.(2023)Osco, Wu, de~Lemos, Gon{\c{c}}alves, Ramos, Li, and Junior]{osco2023segment}
Lucas~Prado Osco, Qiusheng Wu, Eduardo~Lopes de Lemos, Wesley~Nunes Gon{\c{c}}alves, Ana Paula~Marques Ramos, Jonathan Li, and Jos{\'e}~Marcato Junior.
\newblock The segment anything model (sam) for remote sensing applications: From zero to one shot.
\newblock \emph{International Journal of Applied Earth Observation and Geoinformation}, 124:\penalty0 103540, 2023.

\bibitem[Ouyang et~al.(2020)Ouyang, Biffi, Chen, Kart, Qiu, and Rueckert]{ouyang2020self}
Cheng Ouyang, Carlo Biffi, Chen Chen, Turkay Kart, Huaqi Qiu, and Daniel Rueckert.
\newblock Self-supervision with superpixels: Training few-shot medical image segmentation without annotation.
\newblock In \emph{Computer Vision--ECCV 2020: 16th European Conference, Glasgow, UK, August 23--28, 2020, Proceedings, Part XXIX 16}, pages 762--780. Springer, 2020.

\bibitem[Ouyang et~al.(2022)Ouyang, Biffi, Chen, Kart, Qiu, and Rueckert]{ouyang2022self}
Cheng Ouyang, Carlo Biffi, Chen Chen, Turkay Kart, Huaqi Qiu, and Daniel Rueckert.
\newblock Self-supervised learning for few-shot medical image segmentation.
\newblock \emph{IEEE Transactions on Medical Imaging}, 41\penalty0 (7):\penalty0 1837--1848, 2022.

\bibitem[Rakelly et~al.(2018)Rakelly, Shelhamer, Darrell, Efros, and Levine]{rakelly2018conditional}
Kate Rakelly, Evan Shelhamer, Trevor Darrell, Alyosha Efros, and Sergey Levine.
\newblock Conditional networks for few-shot semantic segmentation.
\newblock 2018.

\bibitem[Ravi et~al.(2024)Ravi, Gabeur, Hu, Hu, Ryali, Ma, Khedr, R{\"a}dle, Rolland, Gustafson, et~al.]{ravi2024sam}
Nikhila Ravi, Valentin Gabeur, Yuan-Ting Hu, Ronghang Hu, Chaitanya Ryali, Tengyu Ma, Haitham Khedr, Roman R{\"a}dle, Chloe Rolland, Laura Gustafson, et~al.
\newblock Sam 2: Segment anything in images and videos.
\newblock \emph{arXiv preprint arXiv:2408.00714}, 2024.

\bibitem[Roy et~al.(2020)Roy, Siddiqui, P{\"o}lsterl, Navab, and Wachinger]{roy2020squeeze}
Abhijit~Guha Roy, Shayan Siddiqui, Sebastian P{\"o}lsterl, Nassir Navab, and Christian Wachinger.
\newblock ‘squeeze \& excite’guided few-shot segmentation of volumetric images.
\newblock \emph{Medical image analysis}, 59:\penalty0 101587, 2020.

\bibitem[Shaban et~al.(2017)Shaban, Bansal, Liu, Essa, and Boots]{shaban2017one}
Amirreza Shaban, Shray Bansal, Zhen Liu, Irfan Essa, and Byron Boots.
\newblock One-shot learning for semantic segmentation.
\newblock \emph{arXiv preprint arXiv:1709.03410}, 2017.

\bibitem[Shen et~al.(2023)Shen, Li, Jin, and Liu]{shen2023q}
Qianqian Shen, Yanan Li, Jiyong Jin, and Bin Liu.
\newblock Q-net: Query-informed few-shot medical image segmentation.
\newblock In \emph{Proceedings of SAI Intelligent Systems Conference}, pages 610--628. Springer, 2023.

\bibitem[Shi et~al.(2022)Shi, Wei, Zhang, Lu, Ning, Chen, Ma, and Zheng]{shi2022dense}
Xinyu Shi, Dong Wei, Yu Zhang, Donghuan Lu, Munan Ning, Jiashun Chen, Kai Ma, and Yefeng Zheng.
\newblock Dense cross-query-and-support attention weighted mask aggregation for few-shot segmentation.
\newblock In \emph{European Conference on Computer Vision}, pages 151--168. Springer, 2022.

\bibitem[Simonyan and Zisserman(2014)]{simonyan2014very}
Karen Simonyan and Andrew Zisserman.
\newblock Very deep convolutional networks for large-scale image recognition.
\newblock \emph{arXiv preprint arXiv:1409.1556}, 2014.

\bibitem[Snell et~al.(2017)Snell, Swersky, and Zemel]{snell2017prototypical}
Jake Snell, Kevin Swersky, and Richard Zemel.
\newblock Prototypical networks for few-shot learning.
\newblock \emph{Advances in neural information processing systems}, 30, 2017.

\bibitem[Tang et~al.(2021)Tang, Liu, Sun, Yan, and Xie]{tang2021recurrent}
Hao Tang, Xingwei Liu, Shanlin Sun, Xiangyi Yan, and Xiaohui Xie.
\newblock Recurrent mask refinement for few-shot medical image segmentation.
\newblock In \emph{Proceedings of the IEEE/CVF international conference on computer vision}, pages 3918--3928, 2021.

\bibitem[Tian et~al.(2020)Tian, Zhao, Shu, Yang, Li, and Jia]{tian2020prior}
Zhuotao Tian, Hengshuang Zhao, Michelle Shu, Zhicheng Yang, Ruiyu Li, and Jiaya Jia.
\newblock Prior guided feature enrichment network for few-shot segmentation.
\newblock \emph{IEEE transactions on pattern analysis and machine intelligence}, 44\penalty0 (2):\penalty0 1050--1065, 2020.

\bibitem[Wang et~al.(2024)Wang, Zhang, Du, Xu, Liu, Tao, and Zhang]{wang2024samrs}
Di Wang, Jing Zhang, Bo Du, Minqiang Xu, Lin Liu, Dacheng Tao, and Liangpei Zhang.
\newblock Samrs: Scaling-up remote sensing segmentation dataset with segment anything model.
\newblock \emph{Advances in Neural Information Processing Systems}, 36, 2024.

\bibitem[Wang et~al.(2019)Wang, Liew, Zou, Zhou, and Feng]{wang2019panet}
Kaixin Wang, Jun~Hao Liew, Yingtian Zou, Daquan Zhou, and Jiashi Feng.
\newblock Panet: Few-shot image semantic segmentation with prototype alignment.
\newblock In \emph{proceedings of the IEEE/CVF international conference on computer vision}, pages 9197--9206, 2019.

\bibitem[Wang et~al.(2022)Wang, Zhou, and Zheng]{wang2022few}
Runze Wang, Qin Zhou, and Guoyan Zheng.
\newblock Few-shot medical image segmentation regularized with self-reference and contrastive learning.
\newblock In \emph{International Conference on Medical Image Computing and Computer-Assisted Intervention}, pages 514--523. Springer, 2022.

\bibitem[Wu et~al.(2022)Wu, Xiao, and Liang]{wu2022dual}
Huisi Wu, Fangyan Xiao, and Chongxin Liang.
\newblock Dual contrastive learning with anatomical auxiliary supervision for few-shot medical image segmentation.
\newblock In \emph{European Conference on Computer Vision}, pages 417--434. Springer, 2022.

\bibitem[Wu and Xu(2024)]{wu2024one}
Junde Wu and Min Xu.
\newblock One-prompt to segment all medical images.
\newblock In \emph{Proceedings of the IEEE/CVF Conference on Computer Vision and Pattern Recognition}, pages 11302--11312, 2024.

\bibitem[Wu et~al.(2023)Wu, Ji, Liu, Fu, Xu, Xu, and Jin]{wu2023medical}
Junde Wu, Wei Ji, Yuanpei Liu, Huazhu Fu, Min Xu, Yanwu Xu, and Yueming Jin.
\newblock Medical sam adapter: Adapting segment anything model for medical image segmentation.
\newblock \emph{arXiv preprint arXiv:2304.12620}, 2023.

\bibitem[Zhang et~al.(2021{\natexlab{a}})Zhang, Xiao, and Qin]{zhang2021self}
Bingfeng Zhang, Jimin Xiao, and Terry Qin.
\newblock Self-guided and cross-guided learning for few-shot segmentation.
\newblock In \emph{Proceedings of the IEEE/CVF conference on computer vision and pattern recognition}, pages 8312--8321, 2021{\natexlab{a}}.

\bibitem[Zhang et~al.(2021{\natexlab{b}})Zhang, Kang, Yang, and Wei]{zhang2021few}
Gengwei Zhang, Guoliang Kang, Yi Yang, and Yunchao Wei.
\newblock Few-shot segmentation via cycle-consistent transformer.
\newblock \emph{Advances in Neural Information Processing Systems}, 34:\penalty0 21984--21996, 2021{\natexlab{b}}.

\bibitem[Zhang et~al.(2018)Zhang, Isola, Efros, Shechtman, and Wang]{zhang2018unreasonable}
Richard Zhang, Phillip Isola, Alexei~A Efros, Eli Shechtman, and Oliver Wang.
\newblock The unreasonable effectiveness of deep features as a perceptual metric.
\newblock In \emph{Proceedings of the IEEE conference on computer vision and pattern recognition}, pages 586--595, 2018.

\bibitem[Zhang et~al.(2020)Zhang, Wei, Yang, and Huang]{zhang2020sg}
Xiaolin Zhang, Yunchao Wei, Yi Yang, and Thomas~S Huang.
\newblock Sg-one: Similarity guidance network for one-shot semantic segmentation.
\newblock \emph{IEEE transactions on cybernetics}, 50\penalty0 (9):\penalty0 3855--3865, 2020.

\bibitem[Zhu et~al.(2023)Zhu, Wang, Xin, and Zhang]{zhu2023few}
Yazhou Zhu, Shidong Wang, Tong Xin, and Haofeng Zhang.
\newblock Few-shot medical image segmentation via a region-enhanced prototypical transformer.
\newblock In \emph{International Conference on Medical Image Computing and Computer-Assisted Intervention}, pages 271--280. Springer, 2023.

\bibitem[Zhuang(2018)]{zhuang2018multivariate}
Xiahai Zhuang.
\newblock Multivariate mixture model for myocardial segmentation combining multi-source images.
\newblock \emph{IEEE transactions on pattern analysis and machine intelligence}, 41\penalty0 (12):\penalty0 2933--2946, 2018.

\end{thebibliography}
}


\end{document}